\documentclass[10pt,twocolumn]{article}

\usepackage[T1]{fontenc}
\usepackage{lmodern}
\usepackage[letterpaper,top=0.70in,bottom=0.74in,left=0.70in,right=0.70in]{geometry}
\usepackage{microtype}
\usepackage{amsmath,amssymb}
\usepackage{booktabs}
\usepackage{array}
\usepackage{tabularx}
\usepackage{multirow}
\usepackage{enumitem}
\usepackage{natbib}
\usepackage{xcolor}
\usepackage{hyperref}
\usepackage{url}
\usepackage{tikz}
\usepackage{pgfplots}
\usepackage{balance}
\usepackage{placeins}
\usetikzlibrary{positioning,arrows.meta,calc}
\usepgfplotslibrary{groupplots}
\pgfplotsset{compat=1.18}

\definecolor{StudyBlue}{HTML}{245DA8}
\definecolor{StudyTeal}{HTML}{147B70}
\definecolor{StudyGold}{HTML}{A66A16}
\definecolor{StudyInk}{HTML}{18212B}
\definecolor{StudyMuted}{HTML}{5E6A75}
\definecolor{StudyGrid}{HTML}{D5DCE3}
\definecolor{StudyFill}{HTML}{EFF4F8}

\hypersetup{
  pdftitle={The Prompt Is Not the Query: How Request State Evolves Across Multi-Turn AI Conversations},
  pdfauthor={Benjamin Tannenbaum},
  colorlinks=true,
  linkcolor=StudyBlue,
  citecolor=StudyBlue,
  urlcolor=StudyBlue
}

\setlist{nosep,leftmargin=*}

\newcolumntype{Y}{>{\raggedright\arraybackslash}X}

\newcommand{\commercialN}{670}
\newcommand{\prismN}{7,463}
\newcommand{\crossN}{8,133}

\title{\vspace{-0.45in}\textbf{The Prompt Is Not the Query:}\\
How Request State Evolves Across Multi-Turn AI Conversations}

\author{
  Benjamin Tannenbaum\\
  Aiso, Tel Aviv, Israel\\
  \texttt{ben@getaiso.com}
}

\date{\vspace{-0.12in}}

\begin{document}
\maketitle

\begin{abstract}
AI-search evaluation commonly treats a prompt as a stable query that can be
counted, classified, and replayed in isolation. A conversation makes that unit
of analysis questionable: each user turn can add a constraint, revise an
assumption, request evidence, or refer to alternatives established earlier.
We replace latent ``intent'' with an observable construct,
\emph{conversation-conditioned request state}, and measure how that state is
distributed across user turns. The analysis reuses frozen rules and the
governed cohort of a preceding conversation study: \commercialN{} English
commercial multi-turn conversations in a discovery-replication design and
\prismN{} public PRISM conversations from 1,389 participants. In the
commercial corpus, the final prompt contains a median 35.6\% of the session's
unique user-side content vocabulary; in PRISM, the median is 36.4\%. The final
prompt contains at most half of that vocabulary in 68.4\% and 74.3\% of
conversations, respectively. More importantly, transparent rules detect at
least one request-state dimension in history but not in the final prompt in
50.3\% of commercial conversations and 44.8\% of PRISM conversations. Among
dimension-bearing conversations, the final prompt reproduces the full observed
dimension set in only 26.1\% and 26.2\%. At the same time, the final prompt
adds a previously unseen dimension in 17.9\% and 19.3\%, showing that the
endpoint is neither a summary nor merely a reference: it is often another
state update. Length-matched nulls show that low lexical coverage is largely a
consequence of turn length, so vocabulary results are interpreted as
information availability, not semantic drift. The categorical results support
session-level measurement for AI search. They do not estimate the causal
effect of history on model answers.
\end{abstract}

\section{Introduction}

Keyword search receives a query. A conversational system receives a message
inside an interaction. Measurement practice has not fully absorbed that
difference. Prompt panels, visibility trackers, benchmark suites, and many
internal evaluations still replay isolated messages and treat the answer as a
property of the message alone.

This assumption is convenient, but it can fail even when the broad topic never
changes. A user may ask for options, add a budget after seeing them, reject one
assumption, request evidence, and end with ``Which one would you pick?'' The
final string is locally short because much of its actionable meaning is stored
elsewhere. Conversely, the final turn can introduce a new use case or
comparison criterion that did not appear previously. Calling every turn an
independent ``query'' collapses these distinct operations.

The concept of intent creates a second problem. In keyword retrieval, an intent
label can be a useful summary of one request. In dialogue, ``intent'' may refer
to a latent psychological goal, a topic, a speech act, a task specification, a
current preference, or an annotation category. Those objects need not move
together. Recent evidence that users' self-reported thoughts differ
semantically from their messages makes it especially important not to infer
private mental state from transcripts alone \citep{jin2026thoughttrace}.

We therefore define a narrower transcript-level construct:
\begin{quote}
\emph{Conversation-conditioned request state is the explicit task
specification and discourse dependence needed to interpret the user's current
turn within the observed session.}
\end{quote}
The state can include goals, constraints, alternatives, corrections,
evaluation criteria, and evidence requirements. A turn can contribute a
\emph{delta} to this state without restating it. This distinction yields a
testable question:
\begin{quote}
\emph{How much of the observed request state is contained in any one prompt,
and what does an isolated final prompt leave in history?}
\end{quote}

We answer this question with \crossN{} real human--LLM conversations from two
sources. The proprietary corpus contains a Drive-based discovery cohort and a
non-overlapping replication cohort. PRISM provides public external validation
across values-guided, controversy-guided, and unguided protocols
\citep{kirk2024prism}. We use only user turns. Assistant text is excluded from
all outcomes.

The paper contributes:
\begin{itemize}
  \item a formal distinction between a turn-local prompt and cumulative
  conversation-conditioned request state;
  \item endpoint measures of history-resident state, endpoint-added state, and
  best-single-turn representation;
  \item real results from a commercial discovery-replication design and a
  participant-clustered public validation;
  \item fixed-depth trajectories that show gradual state accumulation within
  the same conversations; and
  \item an explicit boundary between observational request-state measurement
  and the unmeasured causal effect of history on answers.
\end{itemize}

\section{Related Work}

\subsection{Conversational information needs}

Conversational information seeking treats the information need as an
interaction-level object rather than a property of one utterance
\citep{zamani2023conversational}. Conversational query rewriting makes the
dependence operational by converting context-dependent turns into
self-contained retrieval inputs \citep{elgohary2019can,qian2022explicit}.
Later systems select or enhance history before rewriting
\citep{mo2023convgqr,mo2024chiq}. These methods show why context can be useful
to a retriever. They do not estimate how much of a naturally observed session
is represented by its endpoint prompt.

Goal-clarification work questions the assumption that users enter a dialogue
with a fully specified task \citep{liu2022goals}. Classical critiques of
intent-plus-slot models likewise argue that collaborative dialogue requires a
richer dialogue state than an atomic intent label and simple slot fillers
\citep{cohen2019slots}. Transition-aware dialogue research models changes
between conversation modes \citep{yoon2025transition}. Our contribution is
descriptive rather than generative: we measure where explicit request evidence
appears across real user turns without assigning a latent goal.

\subsection{History use by language models}

Perturbation studies have found that neural dialogue systems can be insensitive
to history \citep{sankar2019history}, while newer work shows that language
models can degrade when instructions are disclosed progressively across turns
\citep{laban2025lost}. \citet{huang2026ownwords} distinguish user-authored
history from prior assistant text and show that assistant history may be
unnecessary or harmful in some in-the-wild cases.

Those studies manipulate model input or compare generated answers. The present
paper does neither. It measures whether the user-authored specification itself
is distributed. A paired rerun of each final prompt with and without history
would be required to estimate answer changes. The observational result here is
a prerequisite for such experiments, not a substitute for them.

\subsection{Measurement units in AI search}

Conversational answers can compress multiple conventional query and source
actions into one response \citep{tannenbaum2026density}. The present paper
examines the complementary input-side compression: a short prompt can stand
for a state assembled across prior turns. Together, these perspectives imply
that both the input and output units of conversational search differ from
their keyword-search analogues.

\section{Construct and Questions}

\subsection{A prompt is a state update}

Let a conversation contain user turns $u_1,\ldots,u_T$. Let $V_t$ be the set
of unique lowercased non-stopword content tokens in $u_t$. Let $D_t$ be the set
of explicit request-state dimensions detected in $u_t$, and let
\[
S_t = \bigcup_{i=1}^{t} D_i
\]
be the cumulative observed categorical state through turn $t$.

The final prompt can relate to history in two directions. The
\emph{history-resident set}
\[
H_T = S_{T-1} \setminus D_T
\]
contains dimensions observed earlier but not restated at the endpoint. The
\emph{endpoint-added set}
\[
A_T = D_T \setminus S_{T-1}
\]
contains dimensions first observed at the endpoint. When both are nonempty,
the final prompt simultaneously omits prior evidence and adds new evidence. It
is better understood as a state update than a summary.

This representation is deliberately modest. $S_t$ is not a belief state, a
stable preference, or a claim about what the user ``really'' wants. A cue can
be delayed disclosure, clarification, response-contingent learning, correction,
or genuine preference change. The transcript alone cannot separate those
mechanisms.

\begin{table}[t]
\centering
\small
\begin{tabularx}{\columnwidth}{@{}>{\bfseries}p{0.43in}Y@{}}
\toprule
Turn & Constructed user message and observed state delta\\
\midrule
$u_1$ & ``Help me choose a light laptop for travel.'' Adds use case and
attribute.\\
$u_2$ & ``Under \$900, with good Linux support.'' Adds budget and another
attribute.\\
$u_3$ & ``Not that one. Which is better?'' Adds correction and comparison;
does not restate budget.\\
\bottomrule
\end{tabularx}
\caption{Constructed illustration, not a dataset excerpt. Each prompt changes
the observed request state without reproducing the complete state.}
\label{tab:illustration}
\end{table}

\subsection{Research questions}

\begin{description}[style=nextline,leftmargin=0pt,labelindent=0pt,itemsep=3pt]
\item[RQ1: Endpoint representation.] How much of the session's user-side
content vocabulary appears in the final prompt?
\item[RQ2: Categorical state.] How often does the final prompt leave explicit
dimensions in history, add a new dimension, or do both?
\item[RQ3: Single-prompt adequacy.] Does any individual user turn contain most
of the observed lexical or categorical state?
\item[RQ4: Trajectory.] How do endpoint representation and cumulative state
change with observed conversation depth?
\end{description}

\section{Data}

\subsection{Commercial cohorts}

The commercial source is a proprietary, consent-governed research corpus. The
discovery export was retrieved from Google Drive and contains 1,180 rows.
Identifier deduplication, retaining the richest transcript for repeated IDs,
produces 823 unique records and 809 with at least one parseable user turn.
There are 302 English multi-turn conversations in the discovery analysis.

The non-overlapping replication source contains 1,425 rows from 43 additional
governed CSV exports. After identifier deduplication, exclusion of every ID
present in discovery, and removal of records without a parseable user turn,
940 usable conversations remain. Of these, 368 are English and multi-turn.
The pooled commercial analysis uses \commercialN{} English multi-turn
conversations. The exact pooled construction allows a richer record from an
overlapping identifier to replace its shorter copy, matching the preceding
study's cohort.

Conversation roles are parsed from the transcript body, not from the
convenience opening-prompt field. English is determined from an explicit
language label when present, then by a conservative Latin-character and common
function-word rule when it is absent. Collection names are replaced with
anonymous labels in all outputs.

\subsection{PRISM external validation}

PRISM contains conversations collected from participants across countries,
models, and interaction protocols \citep{kirk2024prism}. We retain
conversations with at least two user turns whose released metadata cover every
user turn, mark every user turn as English, and contain no human-text PII flag.
From 8,011 released conversations, 488 fail the English criterion and 60 have
a human-text PII flag, leaving \prismN{} conversations from 1,389
participants. The primary cohort includes 2,308 values-guided, 2,279
controversy-guided, and 2,876 unguided conversations.

PRISM presents multiple sampled model candidates for user turns rather than
one simple alternating assistant transcript. We therefore use only the human
user turns, as in the preceding analysis. No assistant candidate is treated as
the selected conversation history.

\begin{table*}[t]
\centering
\small
\begin{tabularx}{\textwidth}{@{}lrrrY@{}}
\toprule
Source & Released rows & Usable frame & English multi-turn & Primary role in study\\
\midrule
Commercial discovery & 1,180 & 809 & 302 & Drive-based discovery cohort\\
Commercial replication & 1,425 & 940 non-overlapping & 368 & Later governed replication\\
Commercial pooled & 2,605 & 1,749 unique usable & \commercialN{} & Naturalistic commercial frame\\
PRISM & 8,011 & \prismN{} eligible & \prismN{} & Public cross-domain validation\\
\bottomrule
\end{tabularx}
\caption{Cohort audit. ``Usable'' commercial records require at least one
parsed user turn; deeper outcomes use the English multi-turn subset. PRISM is
protocol-defined as multi-turn and cannot estimate natural continuation.}
\label{tab:data}
\end{table*}

\section{Measures and Inference}

\subsection{Lexical representation}

For each conversation, content vocabulary is the union
$V=\bigcup_{t=1}^{T}V_t$. Final-prompt coverage is
\[
L_T=\frac{|V_T|}{|V|}.
\]
We also record whether $L_T\leq 0.5$, whether $L_T<1$, and the best coverage
achieved by any individual turn,
\[
B=\max_t \frac{|V_t|}{|V|}.
\]
The indicator $B\leq0.5$ identifies sessions in which no one prompt contains
a majority of the observed content types.

Token coverage is length-sensitive. We therefore compute an analytic
length-matched null for the final turn: sample the observed number of final-turn
content-token instances without replacement from all user-side content-token
instances, then calculate the expected share of unique types represented. This
null tests selective lexical retention, not semantic equivalence.

\subsection{Explicit request-state dimensions}

We reuse, without tuning to the present endpoints, nine transparent
case-insensitive cue families from the preceding analysis:
\begin{enumerate}[label=(\arabic*),leftmargin=*,itemsep=1pt]
  \item price or budget;
  \item location or proximity;
  \item persona or use case;
  \item attribute requirement;
  \item time;
  \item alternatives;
  \item correction or redirect;
  \item comparison or evaluation; and
  \item explanation or evidence.
\end{enumerate}
The constraint patterns require explicit terms such as ``budget,'' ``nearby,''
``must,'' or ``weekend.'' Decision-stage patterns require explicit alternatives,
corrections, comparisons, or evidence language. Categories overlap.

Primary categorical outcomes are $H_T\neq\emptyset$, $A_T\neq\emptyset$, and
their intersection. Among conversations with $S_T\neq\emptyset$, final
categorical coverage is $|D_T|/|S_T|$. We also test whether $D_T=S_T$ and
whether any single $D_t$ equals $S_T$.

A conservative final-context cue is present when the final turn begins with an
elliptical or continuation form, contains an explicit continuation,
comparison, or correction phrase, or omits a strict constraint category
detected in earlier user turns. Shortness alone is not counted.

\subsection{Trajectory and depth}

A follow-up turn contains a new-dimension event when
$D_t\setminus S_{t-1}\neq\emptyset$. We record the first such turn and the
number of distinct follow-up turns containing an event. To avoid comparing
different conversations at every time point, a fixed-depth analysis follows
the same conversations with at least five user turns and reports cumulative
lexical coverage and categorical completeness after turns one through five.

\subsection{Intervals and sensitivities}

Commercial proportions use Wilson 95\% confidence intervals; medians use
distribution-free order-statistic intervals. Discovery-replication
proportion differences use Newcombe intervals, while median differences use a
5,000-draw percentile bootstrap. PRISM intervals use 2,000 participant-clustered
bootstrap draws because participants contribute multiple conversations.
Rates are never pooled across commercial and PRISM sources.

Sensitivities include exact user-side transcript deduplication, one
conversation per PRISM participant, PRISM's balanced subset, exclusion of
moderation-flagged PRISM records, explicit-English-only commercial discovery,
commercial openings of at most 500 words, substantive final turns, and
leave-one-commercial-collection-out analyses.

\section{Results}

\subsection{The final prompt is compressed}

Table~\ref{tab:main} reports the main results. The final prompt contains a
median 35.6\% of unique user-side content vocabulary in commercial
conversations (95\% CI: 33.3--38.9) and 36.4\% in PRISM
(participant-clustered 95\% CI: 35.7--37.5). It contains at most half in
68.4\% of commercial conversations (64.7--71.8) and 74.3\% of PRISM
conversations (73.1--75.5).

The lexical result is real but mostly mechanical. The length-matched expected
median is 34.4\% commercially and 35.6\% in PRISM. Median
observed-minus-expected coverage is only 0.4 points in both sources. The
observed final turn is at or above its null expectation in 75.8\% and 81.6\%,
respectively. Final turns are not selectively shedding an unusual amount of
vocabulary; they are short relative to the session. The implication is
information availability, not semantic drift.

\begin{table*}[t]
\centering
\small
\setlength{\tabcolsep}{3.8pt}
\begin{tabular}{@{}lrrrrrr@{}}
\toprule
Source & $n$ & Final lexical & Final $\leq50\%$ & History-only & Final-added &
Final state complete\\
& & median [95\% CI] & \% [95\% CI] & dimension \% [95\% CI] &
dimension \% [95\% CI] & \% [95\% CI]$^\dagger$\\
\midrule
Commercial & 670 & 35.6 [33.3, 38.9] & 68.4 [64.7, 71.8] &
50.3 [46.5, 54.1] & 17.9 [15.2, 21.0] & 26.1 [22.3, 30.3]\\
PRISM & 7,463 & 36.4 [35.7, 37.5] & 74.3 [73.1, 75.5] &
44.8 [43.5, 46.2] & 19.3 [18.4, 20.3] & 26.2 [24.9, 27.5]\\
\bottomrule
\end{tabular}
\caption{Primary endpoint results. Values are percentages except $n$.
PRISM intervals are participant-clustered. $^\dagger$Among conversations with
at least one detected request-state dimension: $n=456$ commercial and
$n=4{,}534$ PRISM.}
\label{tab:main}
\end{table*}

\subsection{Request state remains in history}

Categorical cues show why the endpoint is not simply a compressed restatement.
At least one detected dimension appears before the final prompt but not in it
in 50.3\% of commercial conversations (46.5--54.1) and 44.8\% of PRISM
conversations (43.5--46.2). Among dimension-bearing conversations, the final
prompt reproduces the complete session-level dimension set in only 26.1\%
commercially and 26.2\% in PRISM.

The endpoint also contributes new state. A dimension first appears in the
final prompt in 17.9\% of commercial conversations and 19.3\% of PRISM
conversations. Both directions occur together in 8.7\% (6.8--11.0) and 7.6\%
(7.0--8.2): the final prompt leaves some explicit state in history while
introducing something else. This is the signature of a state update, not a
standalone query or a complete summary.

The conservative final-context cue is present in 46.3\% of commercial
conversations (42.5--50.1) and 33.7\% of PRISM conversations (32.5--34.9).
This measure does not infer whether an isolated model could guess the missing
context. It establishes that the observed final prompt either contains an
explicit dependency marker or fails to restate an earlier strict constraint.

\begin{figure}[t]
\centering
\begin{tikzpicture}
\begin{axis}[
  width=\columnwidth,
  height=2.05in,
  ybar,
  bar width=6pt,
  ymin=0,ymax=80,
  ylabel={Share of conversations (\%)},
  symbolic x coords={FinalHalf,HistoryOnly,FinalAdded,ContextCue},
  xtick=data,
  xticklabels={Final $\leq$ half,History-only,Final-added,Context cue},
  x tick label style={rotate=24,anchor=east,font=\scriptsize},
  ytick={0,20,40,60,80},
  ymajorgrids,
  grid style={StudyGrid},
  axis line style={StudyMuted},
  tick style={StudyMuted},
  legend style={draw=none,font=\scriptsize,at={(0.5,1.03)},anchor=south,legend columns=2},
  nodes near coords,
  nodes near coords style={font=\tiny,/pgf/number format/fixed,/pgf/number format/precision=1},
]
\addplot[fill=StudyBlue,draw=StudyBlue] coordinates {
  (FinalHalf,68.4) (HistoryOnly,50.3) (FinalAdded,17.9) (ContextCue,46.3)
};
\addplot[fill=StudyTeal,draw=StudyTeal] coordinates {
  (FinalHalf,74.3) (HistoryOnly,44.8) (FinalAdded,19.3) (ContextCue,33.7)
};
\legend{Commercial,PRISM}
\end{axis}
\end{tikzpicture}
\caption{Final-prompt outcomes by source. ``Final half'' is lexical;
the remaining outcomes use transparent categorical or discourse cues.}
\label{fig:endpoint}
\end{figure}
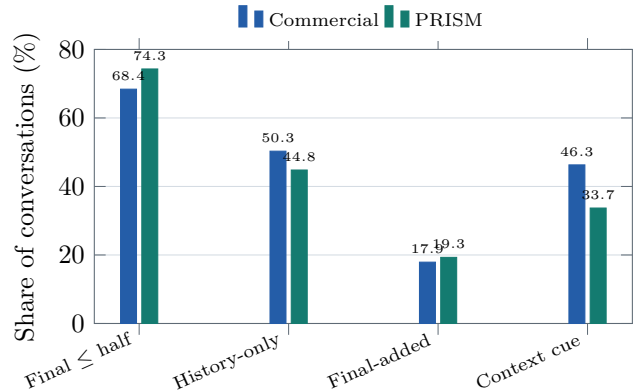

\subsection{No single prompt contains the whole session}

The endpoint is not uniquely weak. In 30.1\% of commercial conversations
(26.8--33.7) and 48.7\% of PRISM conversations (47.2--50.2), no individual
user turn contains a majority of the session's unique content vocabulary.

The same result appears categorically. Among dimension-bearing conversations,
no single user turn contains every detected dimension in 32.2\% commercially
(28.1--36.7) and 24.9\% in PRISM (23.6--26.3). A measurement system that
replays the ``best'' observed prompt would still omit an explicit part of the
session-level state in these cases.

\subsection{State changes gradually with depth}

Across all follow-up transitions, a turn adds at least one previously unseen
dimension in 19.4\% of commercial transitions and 21.2\% of PRISM transitions.
Among conversations with any addition, the first occurs on turn two in 57.5\%
commercially and 58.7\% in PRISM; by turn three, the corresponding cumulative
shares are 77.1\% and 85.1\%. More than one separate follow-up turn adds a new
dimension in 11.0\% of all commercial conversations and 7.1\% of PRISM
conversations. The rules are conservative, so these are demonstrated events,
not estimates of all semantic change.

Depth strengthens the endpoint result. At two user turns, median final lexical
coverage is 54.5\% commercially and 55.6\% in PRISM. At five or more turns, it
falls to 18.2\% and 17.6\%. History-only dimensions rise from 36.5\% to 72.2\%
commercially and from 26.3\% to 71.0\% in PRISM. Conversation depth is an
observed outcome, not an assigned treatment, so these gradients are
descriptive.

\begin{figure*}[t]
\centering
\begin{tikzpicture}
\begin{groupplot}[
  group style={group size=2 by 1,horizontal sep=1.0cm},
  width=0.46\textwidth,
  height=2.25in,
  ymin=0,ymax=80,
  ytick={0,20,40,60,80},
  xlabel={Final observed user-turn depth},
  xtick={1,2,3,4},
  xticklabels={2,3,4,5+},
  ymajorgrids,
  grid style={StudyGrid},
  axis line style={StudyMuted},
  tick style={StudyMuted},
  legend style={draw=none,font=\small,legend columns=2,at={(0.5,1.03)},anchor=south},
]
\nextgroupplot[
  ylabel={Median final lexical coverage (\%)},
  title={Final prompt representation},
]
\addplot[color=StudyBlue,mark=*,thick] coordinates {
  (1,54.5) (2,37.5) (3,29.9) (4,18.2)
};
\addplot[color=StudyTeal,mark=square*,thick] coordinates {
  (1,55.6) (2,36.4) (3,26.3) (4,17.6)
};
\legend{Commercial,PRISM}
\nextgroupplot[
  ylabel={History-only dimension (\%)},
  title={State retained outside final prompt},
]
\addplot[color=StudyBlue,mark=*,thick] coordinates {
  (1,36.5) (2,47.4) (3,51.4) (4,72.2)
};
\addplot[color=StudyTeal,mark=square*,thick] coordinates {
  (1,26.3) (2,44.5) (3,56.5) (4,71.0)
};
\end{groupplot}
\end{tikzpicture}
\caption{Endpoint representation by final conversation depth. Source-specific
curves converge at five or more user turns. Depth is not randomized.}
\label{fig:depth}
\end{figure*}
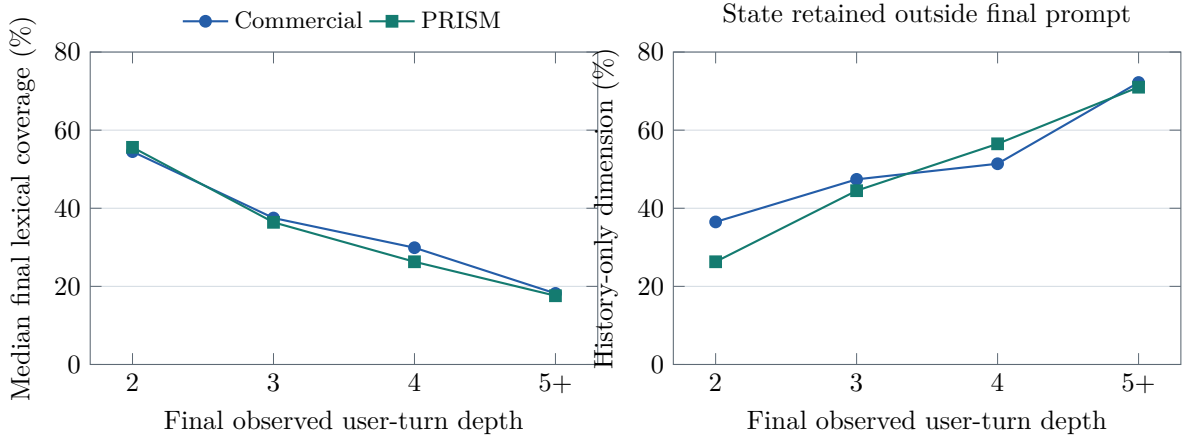

\begin{figure*}[t]
\centering
\begin{tikzpicture}
\begin{groupplot}[
  group style={group size=2 by 1,horizontal sep=1.0cm},
  width=0.46\textwidth,
  height=2.2in,
  ymin=0,ymax=105,
  xmin=1,xmax=5,
  xtick={1,2,3,4,5},
  ytick={0,25,50,75,100},
  xlabel={User turns observed in fixed cohort},
  ymajorgrids,
  grid style={StudyGrid},
  axis line style={StudyMuted},
  tick style={StudyMuted},
  legend style={draw=none,font=\small,legend columns=2,at={(0.5,1.03)},anchor=south},
]
\nextgroupplot[
  ylabel={Median cumulative coverage (\%)},
  title={Content vocabulary},
]
\addplot[color=StudyBlue,mark=*,thick] coordinates {
  (1,24.5) (2,42.9) (3,57.1) (4,73.7) (5,86.8)
};
\addplot[color=StudyTeal,mark=square*,thick] coordinates {
  (1,18.2) (2,37.8) (3,57.1) (4,75.7) (5,100.0)
};
\legend{Commercial,PRISM}
\nextgroupplot[
  ylabel={Categorical state complete (\%)},
  title={Detected request-state dimensions},
]
\addplot[color=StudyBlue,mark=*,thick] coordinates {
  (1,15.7) (2,27.0) (3,40.3) (4,54.7) (5,71.1)
};
\addplot[color=StudyTeal,mark=square*,thick] coordinates {
  (1,12.2) (2,29.1) (3,47.7) (4,65.8) (5,83.5)
};
\end{groupplot}
\end{tikzpicture}
\caption{Cumulative request-state representation within fixed cohorts having
at least five user turns. The lexical panel reports medians. The categorical
panel reports the share whose full observed dimension set has appeared.}
\label{fig:trajectory}
\end{figure*}
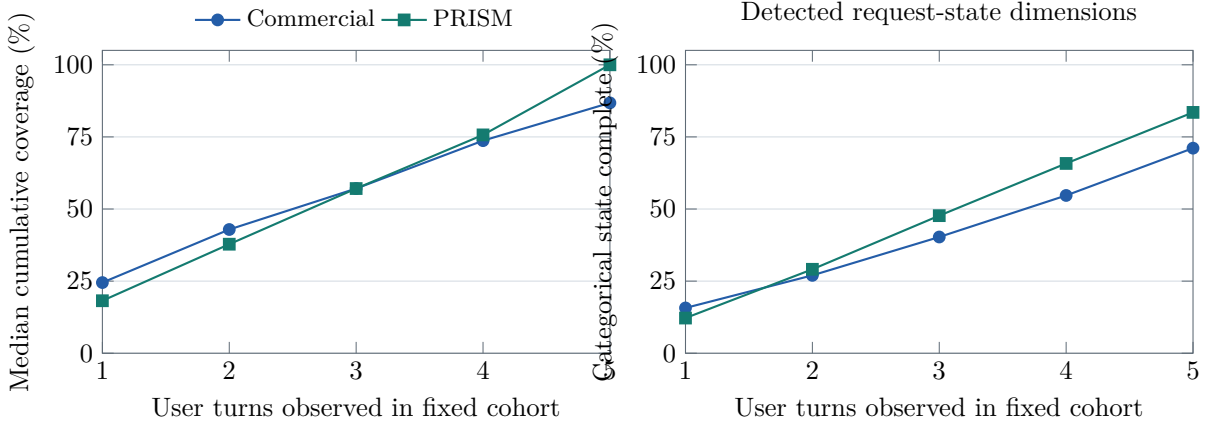

The fixed-depth analysis follows the same 187 commercial and 1,283 PRISM
conversations with at least five user turns. Median cumulative lexical coverage
in the commercial cohort rises from 24.5\% after turn one to 42.9\%, 57.1\%,
73.7\%, and 86.8\% after turns two through five. PRISM rises from 18.2\% to
37.8\%, 57.1\%, 75.7\%, and 100.0\%. Categorical completeness among
dimension-bearing members of these fixed cohorts grows from 15.7\% to 71.1\%
commercially and from 12.2\% to 83.5\% in PRISM. Figure~\ref{fig:trajectory}
shows that request evidence accumulates across several turns rather than
appearing in one predictable location.

\subsection{Discovery, replication, and protocol variation}

The core categorical results replicate closely. A history-only dimension
appears in 52.0\% of discovery and 48.9\% of replication
(replication-minus-discovery difference: -3.1 points, 95\% CI: -10.6 to 4.5).
The final prompt adds a new dimension in 17.5\% and 18.2\% (difference:
0.7 points, -5.3 to 6.4). Conservative endpoint context cues occur in 46.0\%
and 46.5\% (difference: 0.4 points, -7.1 to 8.0).

Lexical endpoint coverage is somewhat higher in replication: 33.3\% versus
37.3\%, a 4.0-point median difference with a wide bootstrap interval
(-1.2 to 13.0). Accordingly, the at-most-half rate falls from 73.8\% to
63.9\% (difference: -10.0 points, -16.8 to -2.9). The lexical difference does
not overturn the categorical replication.

Within PRISM, median final coverage ranges from 35.7\% in values-guided to
37.5\% in controversy-guided conversations. History-only dimensions range
from 42.9\% in controversy-guided and 43.0\% in unguided conversations to
49.0\% in values-guided conversations. Final categorical completeness ranges
from 23.7\% to 27.8\% among dimension-bearing conversations. Protocol affects
the level, but no group approaches endpoint sufficiency.

\section{Implications}

\subsection{The session state is the measurement object}

An isolated prompt panel observes $u_t$. A conversational system responds to
$u_t$ in the presence of a state assembled from earlier turns. The two are not
equivalent measurement objects. The present results suggest three separate
records for AI-search analytics:
\begin{enumerate}
  \item the local user turn;
  \item the cumulative user-authored request state through that turn; and
  \item the supplied assistant history and system context.
\end{enumerate}
Conflating them makes it impossible to tell whether a visibility change came
from a new local instruction, an inherited user constraint, or prior model
framing.

For brand and product measurement, repeated isolated prompts can miss budgets,
use cases, exclusions, and comparisons that were stated earlier. A panel may
therefore estimate an answer distribution for a text string that is not the
unit users actually present to a model. Session-level panels should preserve
turn boundaries, version the history policy, and report which state dimensions
were available at generation time.

\subsection{Implications for evaluation design}

The finding does not imply that all history should always be supplied. History
can contain irrelevant text, model errors, or superseded assumptions
\citep{huang2026ownwords,laban2025lost}. It implies that the history policy is
part of the experimental treatment and must be documented.

A direct answer-effect study should hold the final turn and model fixed, then
compare at least three conditions: full interleaved history, user-turn-only
history, and the isolated final turn. The user-only contrast measures
accumulated user specification; the full-history contrast additionally
captures assistant-established referents and framing. Such an experiment would
answer how much history changes facts, recommendations, brands, sources,
confidence, and constraint satisfaction. The present observational study
shows that this experiment is substantively motivated, but it does not report
its outcome.

\begin{table*}[t]
\centering
\small
\setlength{\tabcolsep}{4pt}
\begin{tabularx}{\textwidth}{@{}Yrrrrr@{}}
\toprule
Sensitivity & $n$ & Final lexical median & Final $\leq50\%$ &
History-only dimension & Final state complete$^\dagger$\\
\midrule
Commercial discovery, explicit English & 163 & 29.9 & 73.0 & 46.6 & 20.0\\
Commercial discovery, opening $\leq500$ words & 284 & 35.1 & 72.2 & 48.9 & 24.5\\
Commercial pooled, substantive endpoint & 584 & 40.0 & 64.4 & 50.5 & 27.7\\
PRISM balanced subset & 6,248 & 35.9 & 74.9 & 45.5 & 25.6\\
PRISM moderation unflagged & 7,125 & 36.8 & 74.1 & 44.2 & 26.5\\
PRISM one conversation per participant & 1,389 & 35.3 & 75.6 & 43.5 & 26.2\\
PRISM exact user-side deduplication & 7,456 & 36.4 & 74.3 & 44.8 & 26.2\\
PRISM substantive endpoint & 6,339 & 40.0 & 70.6 & 45.2 & 27.7\\
\bottomrule
\end{tabularx}
\caption{Sensitivity estimates in percent. A substantive endpoint is not a
pure closure and contains at least three content tokens. $^\dagger$Among
dimension-bearing conversations in each row.}
\label{tab:sensitivity}
\end{table*}

\section{Limitations}

\paragraph{Observable cues are not latent intent.}
The dimension rules detect explicit wording. They have false positives and
false negatives, do not resolve negation or scope perfectly, and cannot infer
unstated preferences. The paper therefore reports detected request-state
dimensions, not true intent. A missing dimension in the final prompt can be
active, superseded, or irrelevant; transcript-only rules cannot decide which.

\paragraph{Lexical coverage is length-dependent.}
The length-matched null nearly reproduces the observed endpoint coverage.
Unique-token coverage should be read as a measure of what text is locally
available, not of topic change or semantic loss.

\paragraph{Conversation depth is endogenous.}
Longer conversations differ from shorter ones in task complexity, protocol,
user behavior, and model behavior. Depth gradients are descriptive. PRISM
requires multiple turns and therefore cannot estimate natural continuation.

\paragraph{Source representativeness is limited.}
The commercial records were selected for governed research relevance and are
not a random sample of global assistant use. PRISM is public and broad but
recruited and protocol-driven. Commercial participant identifiers are
unavailable, so dependence across records cannot be clustered by person.

\paragraph{Assistant influence is unidentified.}
Later user language can be prompted by an assistant's answer. Because the
analysis excludes assistant text and lacks a counterfactual interaction, it
cannot distinguish delayed disclosure from clarification, learning,
preference construction, or model-induced change.

\paragraph{Answer effects are not measured.}
No prompt was rerun. The results show that isolated prompts omit observed
request evidence; they do not show how much a model answer would change when
history is removed.

\FloatBarrier
\section{Robustness and Privacy}

The main estimates remain stable across most sensitivities
(Table~\ref{tab:sensitivity}). Exact PRISM user-side deduplication removes
seven conversations with no rounded change. Selecting one conversation per
participant yields 43.5\% history-only state and 26.2\% final completeness.
The substantive-final-turn sensitivity raises lexical coverage to 40.0\% in
both sources, while history-only dimensions remain at 50.5\% commercial and
45.2\% PRISM. Leave-one-commercial-collection-out estimates range from
49.4\% to 52.0\% for history-only dimensions and from 23.5\% to 27.3\% for
final completeness.

Raw proprietary transcripts remain outside the public output tree. Analysis
artifacts contain aggregates only, no transcript text, IDs, collection names,
or participant IDs. PRISM conversations with any released human-text PII flag
are excluded, survey demographics are unused, and assistant candidates are not
analyzed. Verbatim examples are not published; Table~\ref{tab:illustration} is
constructed.

Conversation logs are intrinsically difficult to anonymize because sparse
language can permit re-identification \citep{narayanan2008robust}. Governance
must therefore rely on access control and minimization, not superficial
redaction alone. The public code records source hashes, filter counts, rule
definitions, and aggregate outputs. Consent-oriented collection systems such
as ShareLM provide a useful model for future public conversation research
\citep{donyehiya2025sharelm}.

\section{Conclusion}

A prompt is a local utterance. A query, in conversational AI, is often a state
distributed across the session. Across \crossN{} real multi-turn
conversations, the final prompt contains roughly one third of observed
user-side content vocabulary. More importantly, nearly half of conversations
retain at least one explicit request-state dimension only in history, and the
final prompt reproduces the complete detected dimension set in only about one
quarter of dimension-bearing sessions. At the same time, roughly one in five
final prompts adds a new dimension.

The endpoint is therefore neither an independent query nor a reliable summary.
It is frequently another delta in an evolving request state. AI-search
measurement should treat the session state and history policy as first-class
parts of the experimental unit. Measuring how that history causally changes
answers is the next experiment.

\balance
\bibliographystyle{plainnat}
\bibliography{references}

@article{zamani2023conversational,
  title = {Conversational Information Seeking},
  author = {Zamani, Hamed and Trippas, Johanne R. and Dalton, Jeff and Radlinski, Filip},
  journal = {Foundations and Trends in Information Retrieval},
  volume = {17},
  number = {3--4},
  pages = {244--456},
  year = {2023},
  doi = {10.1561/1500000081}
}

@inproceedings{elgohary2019can,
  title = {Can You Unpack That? Learning to Rewrite Questions-in-Context},
  author = {Elgohary, Ahmed and Peskov, Denis and Boyd-Graber, Jordan},
  booktitle = {Proceedings of EMNLP-IJCNLP},
  pages = {5918--5924},
  year = {2019},
  doi = {10.18653/v1/D19-1605},
  url = {https://aclanthology.org/D19-1605/}
}

@inproceedings{qian2022explicit,
  title = {Explicit Query Rewriting for Conversational Dense Retrieval},
  author = {Qian, Hongjin and Dou, Zhicheng},
  booktitle = {Proceedings of EMNLP},
  pages = {4725--4737},
  year = {2022},
  doi = {10.18653/v1/2022.emnlp-main.311},
  url = {https://aclanthology.org/2022.emnlp-main.311/}
}

@inproceedings{mo2023convgqr,
  title = {{ConvGQR}: Generative Query Reformulation for Conversational Search},
  author = {Mo, Fengran and Mao, Kelong and Zhu, Yutao and Wu, Yihong and Huang, Kaiyu and Nie, Jian-Yun},
  booktitle = {Proceedings of ACL},
  pages = {4998--5012},
  year = {2023},
  doi = {10.18653/v1/2023.acl-long.274},
  url = {https://aclanthology.org/2023.acl-long.274/}
}

@inproceedings{mo2024chiq,
  title = {{CHIQ}: Contextual History Enhancement for Improving Query Rewriting in Conversational Search},
  author = {Mo, Fengran and Ghaddar, Abbas and Mao, Kelong and Rezagholizadeh, Mehdi and Chen, Boxing and Liu, Qun and Nie, Jian-Yun},
  booktitle = {Proceedings of EMNLP},
  pages = {2253--2268},
  year = {2024},
  doi = {10.18653/v1/2024.emnlp-main.135},
  url = {https://aclanthology.org/2024.emnlp-main.135/}
}

@inproceedings{liu2022goals,
  title = {Where to Go for the Holidays: Towards Mixed-Type Dialogs for Clarification of User Goals},
  author = {Liu, Zeming and Xu, Jun and Lei, Zeyang and Wang, Haifeng and Niu, Zheng-Yu and Wu, Hua},
  booktitle = {Proceedings of ACL},
  pages = {1024--1034},
  year = {2022},
  doi = {10.18653/v1/2022.acl-long.73},
  url = {https://aclanthology.org/2022.acl-long.73/}
}

@inproceedings{cohen2019slots,
  title = {Foundations of Collaborative Task-Oriented Dialogue: What's in a Slot?},
  author = {Cohen, Philip},
  booktitle = {Proceedings of the 20th Annual SIGdial Meeting on Discourse and Dialogue},
  pages = {198--209},
  year = {2019},
  doi = {10.18653/v1/W19-5924},
  url = {https://aclanthology.org/W19-5924/}
}

@article{huang2026ownwords,
  title = {Do {LLM}s Benefit From Their Own Words?},
  author = {Huang, Jenny Y. and Choshen, Leshem and Astudillo, Ramon and Broderick, Tamara and Andreas, Jacob},
  journal = {arXiv preprint arXiv:2602.24287},
  year = {2026},
  doi = {10.48550/arXiv.2602.24287},
  url = {https://arxiv.org/abs/2602.24287}
}

@article{laban2025lost,
  title = {{LLM}s Get Lost In Multi-Turn Conversation},
  author = {Laban, Philippe and Hayashi, Hiroaki and Zhou, Yingbo and Neville, Jennifer},
  journal = {arXiv preprint arXiv:2505.06120},
  year = {2025},
  doi = {10.48550/arXiv.2505.06120},
  url = {https://arxiv.org/abs/2505.06120}
}

@inproceedings{sankar2019history,
  title = {Do Neural Dialog Systems Use the Conversation History Effectively? An Empirical Study},
  author = {Sankar, Chinnadhurai and Subramanian, Sandeep and Pal, Chris and Chandar, Sarath and Bengio, Yoshua},
  booktitle = {Proceedings of ACL},
  pages = {32--37},
  year = {2019},
  doi = {10.18653/v1/P19-1004},
  url = {https://aclanthology.org/P19-1004/}
}

@article{jin2026thoughttrace,
  title = {{ThoughtTrace}: Understanding User Thoughts in Real-World {LLM} Interactions},
  author = {Jin, Chuanyang and Li, Binze and Xie, Haopeng and Fang, Cathy Mengying and Li, Tianjian and Longpre, Shayne and Gu, Hongxiang and Chen, Maximillian and Shu, Tianmin},
  journal = {arXiv preprint arXiv:2605.20087},
  year = {2026},
  doi = {10.48550/arXiv.2605.20087},
  url = {https://arxiv.org/abs/2605.20087}
}

@inproceedings{yoon2025transition,
  title = {Beyond Task-Oriented and Chitchat Dialogues: Proactive and Transition-Aware Conversational Agents},
  author = {Yoon, Yejin and Son, Yuri and So, Namyoung and Kim, Minseo and Cho, Minsoo and Lee, Chanhee and Park, Seungshin and Kim, Taeuk},
  booktitle = {Proceedings of EMNLP},
  year = {2025},
  doi = {10.18653/v1/2025.emnlp-main.672},
  url = {https://aclanthology.org/2025.emnlp-main.672/}
}

@inproceedings{kirk2024prism,
  title = {{PRISM} Alignment Dataset: What Participatory, Representative and Individualised Human Feedback Collectively Reveals},
  author = {Kirk, Hannah Rose and Whitefield, Andrew and Rottger, Paul and Bean, Andrew and Margatina, Katerina and Ciro, Juan and Mosquera, Rafael and Bartolo, Max and Williams, Adina and He, He and Vidgen, Bert and Hale, Scott A.},
  booktitle = {Advances in Neural Information Processing Systems},
  volume = {37},
  year = {2024},
  note = {Datasets and Benchmarks Track},
  doi = {10.52202/079017-3342}
}

@inproceedings{donyehiya2025sharelm,
  title = {The {ShareLM} Collection and Plugin: Contributing Human-Model Chats for the Benefit of the Community},
  author = {Don-Yehiya, Shachar and Choshen, Leshem and Abend, Omri},
  booktitle = {Proceedings of ACL, System Demonstrations},
  pages = {167--177},
  year = {2025},
  doi = {10.18653/v1/2025.acl-demo.17}
}

@inproceedings{narayanan2008robust,
  title = {Robust De-anonymization of Large Sparse Datasets},
  author = {Narayanan, Arvind and Shmatikov, Vitaly},
  booktitle = {IEEE Symposium on Security and Privacy},
  pages = {111--125},
  year = {2008},
  doi = {10.1109/SP.2008.33}
}

@article{tannenbaum2026density,
  title = {Answer-Reconstruction Search Density: Measuring the Query and Source Work Compressed by Conversational Answers},
  author = {Tannenbaum, Benjamin},
  journal = {arXiv preprint arXiv:2607.18904},
  year = {2026},
  doi = {10.48550/arXiv.2607.18904},
  url = {https://arxiv.org/abs/2607.18904}
}

\end{document}